\documentclass[aip,jcp,reprint]{revtex4-1}
\usepackage{amsmath,amsfonts,graphicx,dsfont,epstopdf,amssymb}

\usepackage{xcolor}
\begin{document}


\title{A Skew Dividing Surface for Accurate Nonadiabatic Mean-Field Ring Polymer Rates} 



\author{Britta A. Johnson}
\affiliation{Department of Chemistry and Biochemistry, Cornell University, Ithaca, New York 14850 U.S.A}

\author{Nandini Ananth}
\email[]{ananth@cornell.edu}
\affiliation{Department of Chemistry and Biochemistry, Cornell University, Ithaca, New York 14850 U.S.A}


\date{\today}

\begin{abstract}
    Mean-Field Ring Polymer Molecular Dynamics (MF-RPMD) is a powerful, efficent, and 
    accurate method for approximate quantum dynamic simulations of multi-level system dynamics. 
    Initial efforts to compute nonadiabatic reaction rates using MF-RPMD were not succesful; 
    recent work showed that this can be remedied by including a simple, if {\it adhoc}, 
    correction term that accounts for the formation of `kinked' or mixed electronic state 
    ring polymer configurations. 
    Here, we build on this idea, introducing a electronic state population based reaction 
    coordinate and novel skew dividing 
    surface that constrains nuclear positions to configurations where the reactant and 
    product state potentials are near-degenerate {\it and} that samples kinked electronic state configurations.
    We then demonstrate the numerical accuracy of this method in computing rates for a series
    of nonadiabatic model systems.
\end{abstract}


\maketitle 

\section{Introduction} \label{sec:intro}

Nonadiabatic condensed phase reactions play a critical role in understanding reaction 
mechanisms for a diverse range of interesting systems; these reactions range from 
proton coupled electron transfer in biological systems to charge transfer and 
fluorescence in energetic materials.~\cite{Marcus85,Gray96,Skourtis10,Azzouzi18,Yarkony12,HammesSchiffer08}
The development of accurate and scalable theoretical methods for characterizing nonadiabatic energy and 
charge transfer remains an outstanding challenge. 

The Marcus electron transfer rate is perhaps the most popular of the nonadiabatic rate theories, 
but it is limited by the assumption of parabolic potentials and a classical solvent.~\cite{Marcus} 
The Wolynes rate theory correctly incorporates nuclear quantum effects by using path integral
Monte Carlo methods to compute Fermi Golden Rule rates.~\cite{Wolynes87} Despite its many successes,~\cite{Lawrence18}
this theory does not yield the correct classical rates for high-temperature anharmonic systems.
This failure was remedied by a recently introduced Golden-Rule Quantum Transition State (GR-QTST) ~\cite{Thapa19,Fang19,Fang20} rate 
theory developed based on insights drawn from semiclassical~\cite{Tong20,Sun18,Lawrence19} and ring polymer instanton theories.~\cite{Heller20,Lawrence20_4,Heller20_2,Mattiat18}
Finally, the Linear Golden-Rule approximation for nonadiabatic rates was introduce to address the size
inconsistencies observed in GR-QTST simulations of condensed phase systems.~\cite{Lawrence20_2,Lawrence20_3} These rate theories 
are accurate and efficient but are limited to the golden-rule weak coupling limit. In addition,
while rate theories play a key role in understanding and interpreting experimental studies, and 
can even provide insights into the dominant paths (instantons), the need for detailed mechanistic
insights drives the development of direct dynamic methods for rate calculations.

A range of real-time dynamic methods including  mixed quantum-classical~\cite{Tully90,Kapral06,Jain2015,CrespoOtero18} 
and semiclassical methods\cite{Cao97,Cao95,Miller16,Cotton13,Richardson15,Richardson15_1,Lee16,Ananth16,Church18, Liu18} 
have been used to simulate nonadiabatic reaction dynamics and compute rates; however, many of these methods cannot 
be easily scaled to the simulation of large condensed phase reactions. Path integral
based methods like centroid-molecular dynamics \cite{Cao94,Jan99} and ring polymer molecular dynamics (RPMD) \cite{Craig04} 
have shown particular promise in modeling condensed phase energy transfer 
reactions.\cite{Craig05,Lawrence20,Novikov18,Hele14,Althorpe13,Hele16,Habershon13}
These methods capture nuclear quantum effects like tunneling and zero-point energy while using only
classical trajectories making them suitable for atomistic simulations of charge transfer in condensed phase systems. 
In particular, ring polymer molecular dynamics has been used to accurately calculate thermal rate constants for 
electron transfer (ET) in the normal and activationless regimes, and proton-coupled electron transfer. 
\cite{Menzeleev11,Wilkins15,Kretchmer16a} RPMD has also been extended to systems with coupled electronic states 
with the more successful formulations including mean-field (MF)-RPMD,~\cite{mfnote} 
kinetically constrained (KC)-RPMD,~\cite{Menzeleev14,Kretchmer16b,Kretchmer18}
nonadiabatic RPMD,~\cite{Richardson13} coherent-state RPMD,~\cite{Chowdhury17},
and mapping-variable RPMD.\cite{Ananth10, Ananth13, Duke15, Pierre17} KC-RPMD has been previously used 
to compute reaction rates for a model ET system in the normal and inverted Marcus regimes.
Further, ring polymer surface hopping methods have been developed to add nuclear quantum effects 
to nonadiabatic surface hopping simulations; these methods work well for model systems despite
the fact that the dynamics do not conserve the quantum Boltzmann distribution.~\cite{Lu17,Tao18,Tao19,Ghosh20}
Of the multi-state RPMD methods, MF-RPMD is uniquely efficient, relying on effective state-averaged 
electronic forces to drive nuclear dynamics and requiring no additional variables making it 
suitable for large scale atomistic simulations.

Initial efforts to compute nonadiabatic reaction rates from MF-RPMD 
significantly overestimated the rate.~\cite{Hele_thesis, Kretchmer16b, Duke16}
Previously, one of us showed that this could be remedied by ensuring MF-RPMD 
trajectories sample `kinked' or mixed-electronic state ring polymer configurations
at the dividing surface.~\cite{Duke16} Unfortunately, the {\it adhoc} introduction
of an additional constraint on the types of electronic state configurations sampled
resulted in an inconsistent flux-side expression for the rate constant 
and a difficult-to-implement simulation protocol.
Here, we present a rate expression that is obtained using a novel skew dividing surface
and an electronic population-based reaction coordinate for 
the computation of nonadiabatic rates. We demonstrate the accuracy of 
this approach in  a series of numerical simulations on model nonadiabatic 
ET systems over a wide range of driving forces. 

This paper is organized as follows. In Section \ref{sec:theory} we briefly review the MF-RPMD formalism, and introduce the new skew reaction 
coordinate.  
In section \ref{sec:model_system} we describe the model systems studied here and section \ref{sec:sim_details} outlines 
the details of the rate calculation.
In section \ref{sec:results} we numerically demonstrate sampling by the skew coordinate and present the results of our 
nonadiabatic rate calculations. We summarize our findings in section \ref{sec:conclusion}.

 \section{Theory} \label{sec:theory}

\subsection{Mean-Field Ring Polymer Molecular Dynamics}
In this section we review MF-RPMD \cite{Duke16}. 
For a general $K$-level system with $d$ nuclear degrees of freedom, the diabatic Hamiltonian is 
\begin{equation}
    \hat{H} = \sum_{j=1}^d \frac{ \hat{P}_j^2  }{ 2 M_j} + \sum_{n,m = 1}^K | n \rangle V_{nm}(\hat{R}) \langle m |,
\end{equation}
where $\hat{R}$ and $\hat{P}$ are $d$-dimensional nuclear position and momentum vector operators, respectively. 
The quantum partition function is discretized via repeated insertion of $N$ copies of the 
identity to obtain
\begin{eqnarray}
    \nonumber
    Z &=& Tr[e^{-\beta \hat{H}}] = \int d\{R_\alpha\} \sum_{\{n_\alpha\}=1}^K \prod_{\alpha=1}^N \\
    &\times& 
    \langle R_\alpha,n_\alpha | e^{-\beta_N \hat H} | R_{\alpha+1},n_{\alpha+1} \rangle,
    \label{eq:partfn}
\end{eqnarray}
where $\beta_N = 1/(N k_B T)$, T is temperature, 
N is the number of imaginary time slices (or beads), 
and $R_\alpha$, $n_\alpha$ refer to the nuclear position 
and electronic state of the $\alpha^\text{th}$ bead, respectively.
In Eq.~\ref{eq:partfn}, we use a shorthand for the multi-dimensional 
integral over nuclear coordinates and summation over electronic states, 
$\int d\left\{ R_\alpha \right\}=\int dR_1 \int dR_2..\int dR_N$
and $\sum_{ \{n_\alpha\}=1 }=\sum_{n_1}\sum_{n_2} \ldots \sum_{n_N}$.
Evaluating the matrix elements using the Trotter and short-time approximations, 
we obtain,~\cite{Chandler87,Trotter59}
\begin{eqnarray} \label{eq:PF_PIMD}
    Z \propto \lim_{N \rightarrow \infty} \int \{ dR_{\alpha} \} e^{ -\beta_N V_N(\{R_{\alpha}\})} \text{Tr}[ \Gamma ],
\end{eqnarray}
where
\begin{equation} \label{eq:rp_potential}
    V_N = \sum_{j=1}^d \sum_{\alpha=1}^N \left[\frac{M_j}{2 \beta_n^2} (R_{j,\alpha} - R_{j,\alpha+1})^2 \right],
\end{equation}
\begin{equation}\label{eq:gamma}
    \Gamma = \prod_{\alpha=1}^N M(R_{\alpha}),
\end{equation}
and M is the $K \times K$-dimensional matrix
\begin{equation}\label{eq:m_matrix}
    M_{nm}(R_{\alpha}) = 
    \begin{cases}
    e^{-\beta_N V_{nn}(R_{\alpha})} & n=m \\
    \beta_N V_{nm}(R_{\alpha}) e^{-\beta_N V_{nn}(R_{\alpha})} & n\neq m
    \end{cases}
\end{equation}
Finally, moving the trace in Eq.~\ref{eq:PF_PIMD} into the exponential and introducing $N$ normalized Gaussian integrals in nuclear momenta,
we obtain a phase-space expression for the canonical partition function,
\begin{equation}
    Z \propto \lim_{N \rightarrow \infty} \int \{ dR_{\alpha} \} \int \{ dP_{\alpha} \} e^{ -\beta_N H_N( \{R_{\alpha}\},\{ P_{\alpha} \} ) },
\end{equation}
where the MF-RPMD Hamiltonian is 
\begin{eqnarray}\label{eq:mfrpmd_hamiltonian}
    \nonumber
    H_{N} &=& \sum_{j=1}^{d} \sum_{\alpha=1}^{N} \left[ \frac{M_j}{2 \beta_N^2} (R_{j,\alpha}-R_{j,\alpha+1})^2 + \frac{P_{j,\alpha}^2}{2 M_j} \right]\\
    && \hspace{1in}- \frac{1}{\beta_N} \ln{(\text{Tr}[\Gamma])}.
\end{eqnarray}
The MF-RPMD approximation to quantum real-time thermal correlation functions is obtained 
by sampling initial conditions from an exact quantum canonical ensemble and time-evolving 
trajectories under the MF-RPMD Hamiltonian in Eq.~\ref{eq:mfrpmd_hamiltonian} with $M_j$ chosen
to the physical mass of the nuclei.

\subsection{A Skew Dividing Surface}

For a general reaction with a barrier,
 the rate constant can be written in terms of a flux-side correlation function\cite{Miller83,Chandler87}.
\begin{equation} \label{eq:rate_expression_general}
    k = \lim_{t \to \infty} \frac{ \langle \delta( \xi_{0} - \xi^{\ddagger}) \dot{\xi_0} h(\xi_t - \xi^{\ddagger}) \rangle}
    {\langle h(\xi^{\ddagger} - \xi_0)\rangle},
\end{equation}
where the angular brackets indicate a canonical ensemble average, 
$\delta$ is a delta function, and $h$ is the Heaviside function. 
For a $K$-level system, the generalized reaction coordinate, 
$\xi$, may be a function of the nuclear $\{R\}$ and electronic state $\{n\}$ variables;
$\xi_0$ is the initial value of this coordinate at time $t=0$, 
$\xi_t$ is the value at time $t$, and $\xi^{\ddagger}$ is value at the dividing surface. 

The MF-RPMD Hamiltonian in Eq.~\ref{eq:mfrpmd_hamiltonian} 
is an explicit function of the nuclear positions and momenta, with 
the effective mean-field potential, $\Gamma$, obtained by tracing over all possible
electronic state configurations. Previous attempts to calculate nonadiabatic reaction 
rates using MF-RPMD relied on a nuclear centroid based definition of the dividing 
surface, $\delta\left( \bar R - R^\ddag \right)$, where the reaction coordinate
is the centroid, $\bar R=\frac{1}{N}\sum_\alpha R_\alpha$, and $R^\ddag$ represents the nuclear
configuration at which the two diabatic electronic state potentials cross.
It was shown that MF-RPMD rates computed with this centroid based reaction coordinate 
were accurate for adiabatic systems, but significantly overestimated the rate for 
nonadiabatic systems.~\cite{Hele_thesis, Menzeleev14}
One of us previously showed that the low probability
of sampling `kinked' or multi-electronic state ring polymer configurations for 
nonadiabatic processes, even at the nuclear centroid dividing surface, was responsible
for the failure of MF-RPMD.~\cite{Duke16} Further, it was established that 
a dividing surface obtained by sampling only kinked configurations
{\it and} constraining the nuclear centroid position resulted in accurate
MF-RPMD rates for a range of nonadiabatic model systems.~\cite{Duke16} However,
the mismatch between reaction coordinate and the {\it adhoc} doubly constrained 
dividing surface resulted in a inconsistent flux-side correlation function that was 
challenging to implement in both the normal and inverted Marcus regimes.
Here, we propose an improved MF-RPMD rate expression by introducing 
a new `skew' dividing surface and population-based reaction coordinate. 
We show the skew dividing surface ensures that the reaction bottleneck
is described by kinked ring polymer configurations in the vicinity of 
$R^\ddag$ without any additional constraints.

We start by recognizing that the electronic force on the nuclear degrees
of freedom in MF-RPMD is due to the mean-field potential, 
$\Gamma\left( \left\{ R_\alpha \right\} \right)$ in Eq.~\ref{eq:gamma},
an average over all possible electronic state configurations. 
Because we no longer have explicit electronic state information, 
reaction coordinates, like the centroid coordinate, that work well 
for single surface reactions cannot be used to describe the progress
of a reaction where the reactant and product correspond to distinct 
electronic states. Here, we introduce a coordinate that can 
distinguish between different electronic state configurations
and that uses the relative population changes to track the progress
of the reaction,
\begin{equation}
\Delta P = 
\text{Tr}\left[ \Gamma_0 \right] - \text{Tr}\left[ \Gamma_N \right],
    \label{eq:deltap_def}
\end{equation}
where 
\begin{equation}
    \Gamma_{k} = \prod_{\alpha=1}^{N-k} \left[M(R_\alpha) \mathbb{P}_2 \right]
\prod_{\alpha=N-k+1}^{N} \left[M(R_{\alpha}) \mathbb{P}_1\right],
    \label{eq:gammak}
\end{equation}
and $\mathbb{P}_i$ is the projection operator corresponding to the $i^\text{th}$ electronic
state.
Physically, Eq.~\ref{eq:gammak} is representative of MF-RP configurations where $k$ beads 
are in electronic state 1 and the remaining $N-k$ beads are in state 2.
This reaction coordinate, $\Delta P\left( \left\{ R_\alpha \right\} \right)$, 
like the centroid coordinate, is a function of nuclear bead positions
but unlike the centroid coordinate can distinguish between a system in the reactant
electronic state and product electronic state; $\Delta P$ moves from being 
negative in the reactant region to positive in the product region. 

\begin{figure}[]
    \centering
    \includegraphics[scale=0.7]{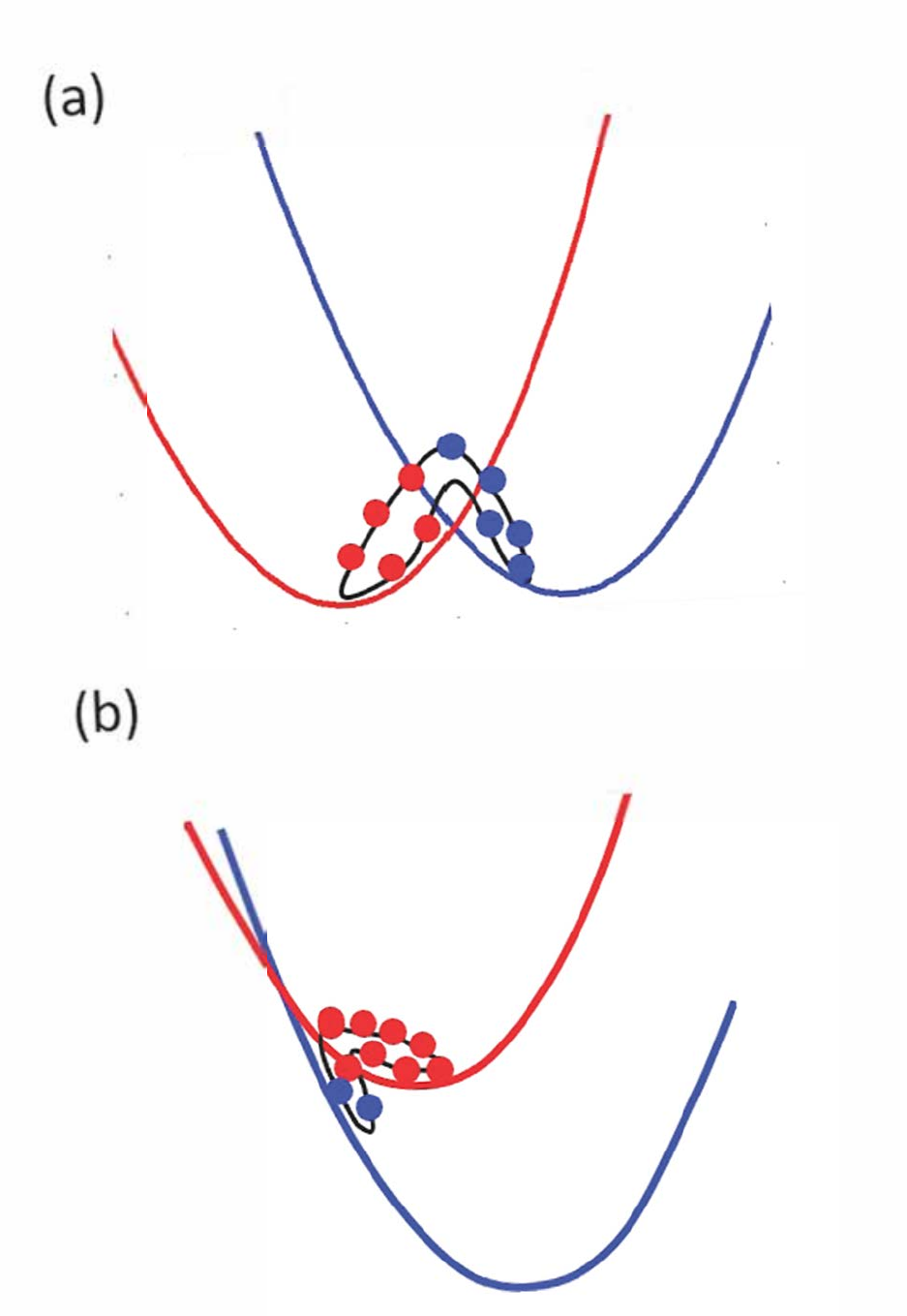}
    \caption{(a) Shows a symmetric two-level system where half the RP beads
        are in the reactant state (red) and the other half are in the product state (blue). 
        As the reactant state is destabilized and the driving force increases, we find that constraining 
nuclear RP configurations to the vicinity of the diabatic crossing can be achieved by 
increasing the number of beads in the reactant state as shown in (b).}
    \label{fig:rp_cartoon}
\end{figure}

In keeping with a reaction that involves a change in electronic state, the 
dividing surface is defined by MF-RP configurations 
sampled on a subset of kinked or mixed electronic-state potentials. 
Constraining the system to this dividing surface is then achieved by 
sampling nuclear configurations 
on $\text{Tr}\left[ \Gamma_{N^\ddag} \right]$ where $0 <  k < N$. 
Values of $N^\ddag$ depend on the driving force; 
as the driving force increases, $N^\ddag$ also increases as shown 
in the cartoon Fig.~\ref{fig:rp_cartoon}.
We note that this idea is in keeping with studies of the MF-RP instanton 
for multistate systems that has an increasing number 
of RP beads in the reactant state as we increase the driving force 
in the Marcus normal regime.~\cite{Ranya20} Formally, the calculated rate
will be independent of the choice of $N^\ddag$, however, 
certain choices can make the simulation
numerically unfeasible.

To find `good' $N^\ddag$ values, we generate histograms of the nuclear configurations
sampled from the kinked MF-RP potential, $\text{Tr}[\Gamma_N^\ddag]$. For simple
system-bath models, we find that $N^\ddag$ values with distributions centered 
around the nuclear configurations at which the two diabatic potentials cross perform 
the best. This can be easily understood: when we sample nuclear configurations 
away from the crossing, in particular towards the reactant minimum, there is a large
barrier to the centroid reaching the crossing, making this a rare event and one that is 
unlikely to be sampled by dynamics on a reasonable timescale. 
This rare event problem is well understood and one that is best avoided by a good 
choice of $N^\ddag$. 
As we demonstrate, for the models considered here, and we expect
for more general model systems, it is not challenging to identify a range of suitable 
$N^\ddag$ values.
%

Using our new reaction coordinate and dividing surface definitions, we obtain an expression
for the rate defined as the long-time limit of a flux-side correlation function,
\begin{equation}\label{eq:rate3}
    k = \lim_{t \rightarrow \infty} 
    \frac{\int d\{R_{\alpha},P_\alpha\} e^{-\beta_N H_{N}^0}
            \text{Tr}\left[ \Gamma_{N^\ddag} \right]
        \dot{\Delta P(0)} h(\Delta P(t))}
        { \int d\{R_{\alpha},P_\alpha\} e^{-\beta_N H_{N}^0}\text{Tr}\left[ \Gamma_N(0) \right] },
\end{equation}
where $H_N^0$ is the free ring polymer Hamiltonian 
that includes only terms on the first line of Eq.~\ref{eq:mfrpmd_hamiltonian}

\subsection{Nonadiabatic Rate Theories}

The Marcus theory (MT) rate for a 
nonadiabatic electron transfer reaction 
with a classical solvent is,
\cite{Marcus1985}
\begin{equation}
k_{\text{MT}} = 
\frac{2 \pi}{\hbar} {| V_{nm}|}^2 
\sqrt{\frac{\beta}{4 \pi \lambda}} 
e^{-\beta \left( \lambda - \epsilon \right)^2 / 4 \lambda},
\label{eq:kmt}
\end{equation}
where $\lambda$ is the reorganization energy,
$\epsilon$ is the driving force, and $V_{nm}$ is
the diabatic coupling between the reactant and product 
electronic states.

Fermi's golden rule rate theory for a nonadiabatic 
electron transfer system where the reactant 
and product state potential energy surfaces are displaced harmonic 
oscillators with frequency $\omega_s$ and with a quantized 
solvent take the simple analytical form\cite{Ulstrup1975,Ulstrup1979}
\begin{equation}
k_{\text{FGR}} = 
\frac{2 \pi}{\hbar \omega_s} {| \Delta|}^2 
e^{v z - S \; \text{coth}\left( z \right)} I_v \left(S \; \text{csch}\left(z\right)\right),
\label{eq:kfgr}
\end{equation}
where $z = \beta \omega_s / 2$, $v = \epsilon/\omega_s$, 
$S = M_s \omega_s V_d^2 / 2 \hbar$,
$M_S$ is the solvent mass, $I_v$ is a modified Bessel function of the first kind, and
$V_d$ is the horizontal displacement of the diabatic potential energy surfaces.

\section{Model System}\label{sec:model_system}
 We calculate the rates for a model condensed-phase ET system with a potential
 \begin{equation}
     V(\hat{R}) = V_S (\hat{s}) + V_B(\hat{R})
 \end{equation}
 where the configuration vector $\hat{R} = \{\hat s,\hat Q\}$ represents the solvent polarization coordinate, $s$, and the bath coordinates, $Q$. 
 The diabatic potential energy matrix is 
 \begin{equation}
     V_S(\hat{s}) = \left( \begin{array}{cc}
             V_{11}(\hat s)
    &  \Delta \\
    \Delta      & 
             V_{22}(\hat s)
     \end{array} \right),
 \end{equation}
 where the diagonal elements are 
 $V_{11}(\hat s)=A \hat{s}^2 + B \hat{s} + \varepsilon$, $V_{22}(\hat s) = A \hat{s}^2 - B \hat{s}$,
 the driving force is represented by $\varepsilon$, and the diabatic coupling a constant, $\Delta$.
 The solvent coordinate is linearly coupled to a thermal bath of $f$ harmonic oscillators,
 \begin{equation}
     V_B(\hat{R}) = \sum_{j=1}^{f} \left[ \frac{1}{2} M_B \omega_j^2 \left( \hat{Q}_j - \frac{c_j \hat{s}}{M_B \omega_j^2} \right)^2 \right]
 \end{equation}
 where $M_S$ and $M_B$ are the solvent and bath mass respectively. 
 The bath is described by an Ohmic spectral
 density
 \begin{equation}
     J(\omega) = \eta \omega e^{-\frac{\omega}{\omega_c}}
 \end{equation}
 where $\omega_c$ is the cutoff frequency and $\eta$ is the dimensionless friction coefficient. 
 The spectral density is discretized into $f$ oscillators \cite{Craig05}
 \begin{equation}
     \omega_j = -\omega_c \ln\left( \frac{j-0.5}{f} \right)
 \end{equation}
 with coupled strengths
 \begin{equation}
     c_j = \omega_j \left( \frac{2 \eta M_B \omega_c}{f \pi} \right)^{1/2}
 \end{equation}
 The ET model parameters are shown in Table \ref{tab:ET_parameters}.
 
 \begin{table}[]
     \centering
     \begin{tabular}{|c|c|}
     \hline
        Parameters   & Value \\
        \hline
        A  & $4.772 \times 10^{-3}$ \\
        B & $2.288 \times 10^{-2}$ \\
        $\varepsilon$ & $0.0 - 0.2366$ \\
        $\Delta$ & $6.69 \times 10^{-7}$ \\
        $M_S$ & $1836.0$ \\
        $M_B$ & $1836.0$ \\
        $f$ & $12$ \\
        $\omega_c$ & $2.28 \times 10^{-3}$ \\
        $\eta / M_{B} \omega_c$ & 1.0 \\
        T & $300$ K \\
        \hline
     \end{tabular}
     \caption{ET model parameters given in atomic units unless otherwise indicated.}
     \label{tab:ET_parameters}
 \end{table}

 \section{Simulation Details}\label{sec:sim_details}
The rate expression in Eq.~\ref{eq:rate3} may still be challenging to 
implement since the numerator requires sampling an ensemble 
constrained to our skew dividing surface while the denominator 
requires efficient sampling of the reactant region. 
To ensure proper sampling of all important regions of 
configuration space, we introduce an identity in
the form of an integral over all possible nuclear RP centroid
configurations to obtain 
\begin{eqnarray}
    k    = 
        \lim_{t \rightarrow \infty} \frac{ \int ds^\prime
        \langle \Gamma_{N^\ddag} \dot{\Delta P}(0) h(\Delta P(t)) \rangle_w }
        {\int ds^\prime \langle \Gamma_{N} \rangle_w}
    \label{eq:rate3}
\end{eqnarray}
where $\langle ... \rangle_w$ 
is used to indicate a phase space ensemble average over the nuclear bead configurations obtained 
by importance sampling from the distribution
\begin{equation}
    w = e^{-\beta_N H_N^0\left( \left\{ R_\alpha, P_\alpha \right\} \right)} \delta(\bar s - s^\prime).
\end{equation}
The numerator and denominator are evaluated using 
a standard Metropolis algorithm to sample free RP configurations
from $e^{-\beta_N H_N^0\left( \left\{ R_\alpha,P_\alpha \right\} \right)}$ in each
window. We then impose the constraint by shifting the solvent RP centroid 
to $\bar{s}$ to the $s^\prime$  value associated with each window.
By scrolling through all possible nuclear RP centroid configurations, we ensure
that the numerator and the denominator are sampled adequately. The integral over 
$s^\prime$ is evaluated using the trapezoid rule.

We establish the mean-field path integral converges with $N=32$ beads 
for all simulations presented here. Importance sampling is performed in each window 
using $~11000$ decorrelated Monte Carlo steps, and 
the final $10000$  configurations are used as initial conditions 
for trajectories evolved under the 
MF-RPMD Hamiltonian in equation \ref{eq:mfrpmd_hamiltonian} with a timestep 
of $0.05$ a.u.
The average initial velocity, $\dot{\Delta P}(0)$ is obtained
by averaging over the finite difference derivative of $\Delta P$ 
calculated for three small intervals of time, $\Delta t= 5, 7, 10$ a.u.
The integral over the solvent centroid configurations in performed over 
150 windows evenly spaced between $s=-4.5$ and $s=+1.5$ for Models I-VIII. 
Model IX simulations are performed with 150 evenly spaced points
between $s=-6.5$ and $s=-0.5$.

\subsubsection{A Modified Implementation in the Inverted Regime}
Physically, the probability of forming kinked configurations, where neighboring beads 
of the ring polymer are in different electronic states, depends on both the magnitude of 
the off-diagonal diabatic coupling and the energy gap between the reactant and product states. 
In the normal regime, we find that the kink probabilities computed using the MF-RP 
potential, $\Gamma\left( \left\{ R_\alpha \right\} \right)$ in Eq.~\ref{eq:gamma} 
with the interaction matrix defined in Eq.~\ref{eq:m_matrix}, do indeed show a 
decreased probability at nuclear configurations where the energetic gap between 
reactant and product states is large. 
However, in the inverted regime, we see a breakdown of this: specifically, we find 
that in regions where the product state is much more favorable than the reactant 
state, the Boltzmann weight of beads in the product state is numerically 
larger than the penalty associated with kink formation, resulting in an 
unphysically large probability of kink formation in regions where the reactant
and product are energetically very different.

We correct for this by a simple modification of the nuclear interaction 
matrix in Eq.~\ref{eq:m_matrix} that is used in computing $\Gamma_N^\ddag$. 
Specifically, we replace $V_{22}$ by $V_{11} + |V_{22}-V_{11}|$ in the appropriate 
off-diagonal term of the interaction matrix. This ensures that the energetic
penalty associated with kink formation is correctly captured, since the value
now depends only on the magnitude of the energy gap between states. Note that this 
is only used in the calculation of $\Gamma_{N^\ddag}$; dynamics are performed using
the MF-RPMD Hamiltonian and the remaining terms in Eq.~\ref{eq:rate3} are unaffected by this change.
 
 \section{Results and Discussion} \label{sec:results}
\begin{figure}[]
     \centering
     \includegraphics[scale=0.78]{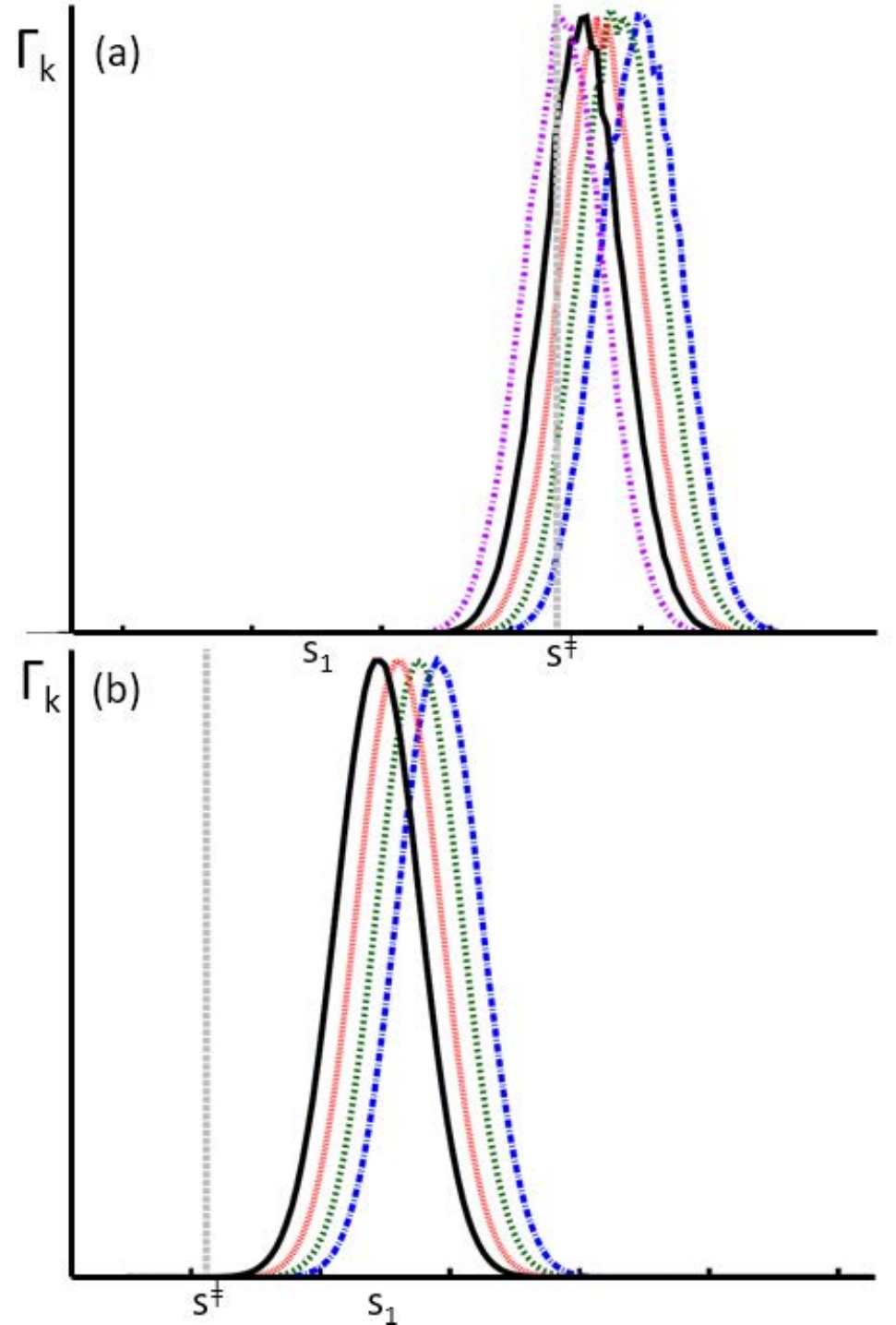}
     \caption{Plots of the $\Gamma_{k}$ functions for (a) model III and (b) model VIII. 
         In both figures, the location of the diabatic crossing is designated by a dotted line at $s^{\ddag}$,
         the reactant minimum indicated by $s_1$, and $\Gamma_k$ with $k=N^\ddag$ is plotted as a black solid line. 
         (a) For model III, $\Gamma_k$ with $k=16$ is shown in blue, $k=17$ is in dark green, $k=18$ is in red,
         $k=19$ is in black, and $k=20$ is in magenta. 
         (b) For model VIII, $\Gamma_k$ with $k=28$ is shown in blue, $k=29$ in dark green, $k=30$ in red, 
     and $k=31$ in black.}
     \label{fig:n1_value_plots}
 \end{figure}
 
 We present the results of our rate calculations for nine model systems
 that differ only in the driving force, $\varepsilon$, values; 
six of these systems (Models I-VI) are located in the normal regime 
and three (Models VII-IX) are in the inverted regime. 
To select the $N^\ddag$ values for each model, we look at the values for 
$\Gamma_{k}\left( \left\{ R_\alpha \right\} \right)$ for individual beads 
as a function of nuclear position.

Figure \ref{fig:n1_value_plots} shows a sample of the $\Gamma_{k}$ curves for 
a model in the normal regime (Models III) and one inverted regime model (Model VIII).
For each model, we find that there is a range of $k$ values where $\Gamma_k$ is maximized 
near the crossing (denoted $s^\ddag$ in the figure), the point where the reactant and 
product state are degenerate. We note that the inverted regime $\Gamma_k$ is modified 
as described in the simulation details; we find that $N^\ddag=31$ is necessary to 
ensure that the dividing surface includes nuclear configurations to the left of 
the reactant minimum towards the diabatic crossing.
For each model in the normal regime, we select an $N^\ddag$ value such that $\Gamma_{N^\ddag}$ 
peaks close to the crossing; the specific values we use are listed in Table \ref{tab:ET_rates}.

 \begin{figure}[h]
     \centering
     \includegraphics[scale=1]{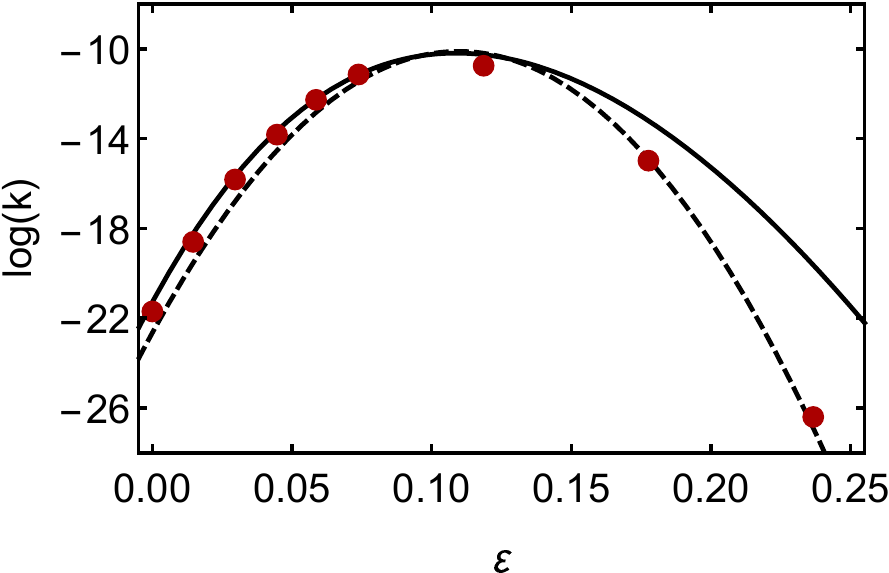}
 \caption{ET rates for models I-IX. MF-RPMD rates with the skew reaction coordinate are shown as red circles, 
 error bars are within the symbol size. The Fermi's Golden rule rates lie along the solid black line and the Marcus 
 theory rates lie along the dashed line}
     \label{fig:et_rates}
 \end{figure}

 We present the MF-RPMD rate results in Fig.~\ref{fig:et_rates} and 
tabulate the corresponding values in Table.~\ref{tab:ET_rates}. We find that the 
new reaction coordinate performs remarkably well in the normal regime, yielding
results that are in quantitative agreement with Fermi Golden Rule rates for all 
six model systems. In the inverted regime, we find good agreement with Marcus theory
rates rather than the golden rule rates. This initially surprising result can be attributed
to the nature of our approximate MF-RPMD dynamics. Decomposing contributions to
the reaction rate from different nuclear configurations, we find that the dominant contribution
is at the crossing. At nuclear configurations where one may reasonably expect tunneling effects 
to allow for a non-zero rate contribution, we find that the Boltzmann-weighted MF-RPMD dynamics
lead to very small values of velocity, $\dot{\Delta P} \approx 0$. In order to obtain FGR rates, then, it 
is likely we will need to move beyond MF-RPMD dynamics.

 \begin{table}[h]
     \centering
     \begin{tabular}{|c|c|c|c|c|c|}
     \hline
     Model  &  $\varepsilon$ &$N^{\ddag}$ & $\log(k_{\text{MT}})$ & $\log(k_{\text{MF})}$ & $\log(k_{\text{FGR}})$\\
        \hline
        I & 0.00 & 16 & $-22.65$ & -21.7 & -21.28 \\
        II & 0.0146 & 17 & $-19.53$ & -18.6 & -18.23 \\
        III & 0.0296 & 19 & $-16.79$ & -15.82 & -15.66 \\
        IV & 0.0446 & 21 & $-14.52 $ & -13.82 & -13.65 \\
         V & 0.0586 & 23 & $-12.83$ & -12.27 & -12.23 \\
         VI & 0.0738 & 25 & $-11.45$ & -11.14 & -11.15 \\
         VII & 0.1186 & 29 & $-10.19$ & -10.75 & -10.26 \\
         VIII & 0.1776 & 31 & $-14.91$ & -14.98 & -13.20 \\
         IX & 0.2366 & 31 & $-26.89$ & -26.4 & -19.63 \\
     \hline
     \end{tabular}
     \caption{ET rates for a range of driving forces. We report the $N^\ddag$ value used for each model, 
         the corresponding Marcus theory rates ($k_\text{MT}$), MF-RPMD rates ($k_\text{MF}$), and 
         Fermi Golden Rule rates ($k_\text{FGR}$) for each model. We see that the MF-RPMD rates are in 
         near-perfect agreement with FGR rates in the normal regime and agree equally well with MT rates in 
         the inverted regimes. All rate constants are in atomic units}
     \label{tab:ET_rates}
 \end{table}

 For four model systems (I and III in the normal regime, and VII and IX in the inverted regime) we show, in Table.~\ref{tab:ndag_dep}, 
 the range of $N^\ddag$ values for which the rate is relatively unchanged. In the normal regime, for large $N^\ddag$ values,
 we find that the dynamic trajectories do not scale the energetic
 barrier necessary to reach the crossing in the timescale of the simulation. 
 In the inverted regime, we have a different problem: small $N^\ddag$ values, corresponding to a large number of 
 beads in the product state, result in an initial distribution of nuclear configurations far from the crossing 
 (indeed, typically we see distributions that peak at configurations between reactant and product minima). 
 Since our dynamics are classical, MF-RPMD trajectories initialized to such configurations do not pass through 
 the reaction bottleneck yielding rates that are significantly higher than expected.

 \begin{table}[h]
     \centering
     \begin{tabular}{|c|c|c|c|c|}
     \hline
     Model  &  $N^\ddag$ range & $\text{log}\left(\bar k\right)$ & $\sigma_k$ & $k_\text{FGR}$ \\
        \hline
        I& 1-21 & -21.63 & 0.3 & -21.28 \\
        I& 11-20 & -21.45 & 0.2 & -21.28 \\
        III & 1-21 & -15.46 & 0.3 & -15.66 \\
        III & 17-21 & -15.61 & 0.3 & -15.66\\
        VII & 27-31 & -10.39 & 0.3 & -10.26 \\
        IX & 29-31 & -26.01 & 0.1 & -19.63 \\
     \hline
     \end{tabular}
     \caption{The rates for models I, III, VII, and IX are computed by 
         averaging over a specific range of $N^\ddag$ values. The reported 
         log average rate, $\text{log}\left( \bar k \right)$,  and 
         standard deviation, $\sigma_k$, indicate the relative robustness
         of our rate with respect to a subset of $N^\ddag$ choices. The 
         options are considerably smaller in the inverted regime where 
         a large number of beads on state 1 are required to ensure that 
         we are not primarily sampling nuclear configurations at the 
 reactant minimum. All rate constants are in atomic units}
     \label{tab:ndag_dep}
 \end{table}

 \section{Conclusion} \label{sec:conclusion}
 We demonstrate that the skew dividing surface and population-based reaction coordinate 
 introduced here can be used to obtain a rigorous MF-RPMD rate theory that is quantitatively accurate for 
 the computation of nonadiabatic reaction rates in a wide range of model systems.
 While we do not compute rates for the adiabatic regime here, we do note that we have 
 previously demonstrated that for model 1, with $N^\ddag=16$, we are able to recover Kramer's rate 
 theory rates for adiabatic reactions~\cite{Duke16}. Future studies will include quantifying 
 the accuracy of our approach over a wider range of models including different friction 
 regimes, temperatures, systems that exhibit multiple transitions states as well as regions of extended 
 electronic coupling. 
 
 MF-RPMD is the most efficient and easy to implement of the RPMD-based methods
 developed to simulate multi-state system dynamics, and we have shown that the 
 use of a skew dividing surface enables the simulation of nonadiabatic processes.
 The form of our dividing surface requires only knowledge of the driving force regime 
 for a particular reaction (information that is typically
 known even for complex systems), placing atomistic simulations of nonadiabatic
 reactions in the condensed phase within reach. 
 
 \section{Acknowledgements}
 The authors would like to acknowledge Prof. Greg Ezra for helpful
 discussion and thank the reviewers of this paper for insightful comments.
 This work was primarily supported by the U.S. Department of Energy,
Office of Basic Energy Sciences, Division of Chemical Sciences,
Geosciences and Biosciences through the Nanoporous Materials
Genome Center under award numbers DE-FG02-17ER16362. Additionally, N. A. 
 acknowledges support from the National Science Foundation Career Award Number CHE1555205.
 
 \section{Data Availability}
 The data that support the findings of this study are available from the corresponding author upon reasonable request.
 

%
%

%


\bibliography{mfrpmd_bib}

\end{document}